\documentclass[intlimits,twoside,a4paper]{article}

\usepackage{amsmath,amssymb}
\usepackage{graphicx}
\usepackage{wrapfig}

\usepackage[T2A]{fontenc}
\usepackage[cp1251]{inputenc}

\usepackage{color}

\usepackage{cmpj2}

\issue{2012}{15}{1}{13604}

\doinumber{10.5488/CMP.15.13604}

\title[Neutron reflectometry in the presence of smooth inter-layer
interfaces]{Investigating the effects of smoothness of interfaces\\ on stability of probing nano-scale thin films \\ by neutron reflectometry}

\author[S.S. Jahromi, S.F. Masoudi]{S.S. Jahromi\thanks{E-mail: s.jahromi@dena.kntu.ac.ir}\,, \ S.F. Masoudi}

\address{Department of Physics, K.N. Toosi University of Technology, P.O. Box 15875-4416, Tehran, Iran}

\authorcopyright{S.S. Jahromi, S.F. Masoudi, 2012}

\date{Received August 5, 2011, in final form February 7, 2012}

\begin{document}

\maketitle

\begin{abstract}
Most of the reflectometry methods which are used for determining the phase of complex reflection coefficient such as reference method
and Variation of Surroundings medium are based on solving the Schr\"{o}dinger equation using a discontinuous and step-like scattering optical potential. However, during the deposition process for making a real sample the two adjacent layers are mixed together and the interface would  not be discontinuous and sharp. The smearing of adjacent layers at the interface (smoothness of interface), would affect the reflectivity, phase of reflection coefficient and reconstruction of the scattering length density (SLD) of the sample. In this paper, we have investigated the stability of reference method in the presence of smooth interfaces. The smoothness of interfaces is considered by using a continuous function scattering potential.  We have also proposed a method to achieve the most reliable output result while retrieving the SLD of the sample.
\keywords neutron reflectometry, thin films, smooth potential, reference method
\pacs 68.35.-p, 63.22.Np
\end{abstract}

\section{Introduction}
In the past decades, neutron reflectometry has been developed as an atomic-scale probe with applications to the study of surface
structure ranging from liquid surfaces to the solid thin films. The type and thickness of the nanostructure materials can be determined
 from measuring the number of neutrons, reflected elastically and specularly from the unknown sample~\cite{Majkrzak1,Xiao-Lin2,Penfold3}.

Neutron reflectometry problems are generally divided into two groups; ``Direct problems'' and ``Inverse problems''. The goal of the direct
 problems is to solve the Schr\"{o}dinger equation for a distinct sample so that the neutron wave function is determined; On the other hand, the application
 of inverse problems is to extract the information of the interacting potential by using the complex reflection coefficient~\cite{Majkrzak1}.

Measuring the intensity of reflected neutrons in terms of the perpendicular component of neutron wave vector to the surface, $q$, provides us with
useful information about the scattering length density of the sample along its depth. As with any scattering technique in which
 only the intensities are measured, the loss of the phase of reflection would result in unresolvable ambiguities on retrieving
the SLD from measurement. Without the knowledge of the phase of reflection, more than one SLD can be found for the same reflectivity
 data. By knowing the phase of the reflection, a unique result would be obtained for the depth profile of the unknown sample~\cite{Kasper4, Masoudi8}.

In the recent years, several methods have been worked out for determining the complex reflection coefficient such as: Dwell time method,
 variation of surroundings medium and the reference layer method which seems to be the most practical~\cite{Majkrzak1}.
\begin{figure}\label{fig1}
\centering{\includegraphics[width=8.8cm]{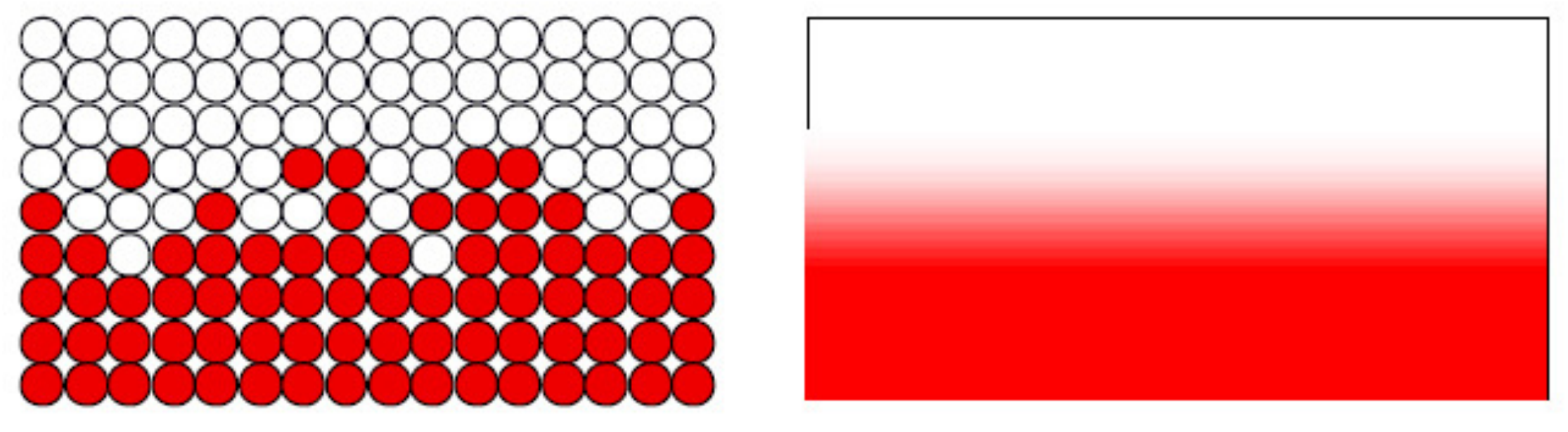}}\\
\centering{ (a) \qquad\qquad\qquad\qquad\qquad\qquad (b)}
\caption{Smearing of the SLD of two adjacent layers at the interface.
 The SLD of layer have been mixed together during the deposition process. (a) Smearing of layers in atomic scale. (b) Continuous variation
 of SLD at the interface in one dimension.}
\end{figure}

 All of these methods are worked out for an ideal sample in which the interface of two adjacent layers is discontinuous and sharp.
 In this case, the reflection coefficient is determined by the solution of one dimensional Schr\"{o}dinger equation for a step-like optical
 potential. However, as we know from a real sample, there is some smearing at the boundaries. This generally happens during the deposition
 of a top layer which is miscible with the bottom material (figure~\ref{fig1}). This process is strongly temperature dependent.
 In this case, the interface is defined as a thin layer across which, the SLD of the sample varies smoothly around the mean position of the interface~\cite{jahromi}.

In most of the simulation methods, the reflectivity measurement and phase determination process is performed for an ideal sample.
 Considering a continuous and smooth varying optical potential at the interface, it would affect the retrieved reflection coefficient.
 In the present work, we have studied the effects of potential smoothness on retrieving the depth profile of the sample and the phase
 of the reflection by considering a continuous potential at boundaries, using reference layers method. As retrieving the SLD from the
 real and imaginary part of the reflection coefficient is very sensitive, we are going to investigate the way one can find the characteristics
 of an unknown sample in the presence of the smoothness effects.

\section{Theory and method}\label{Theory}
Interaction of neutron with nuclei is demonstrated by the neutron optical potential $v(z)$,
which is a function of the scattering length density of the sample,  $v(z)=2\pi\hbar^2\rho(z)/4\pi$
where $\rho(z)$ is the scattering length density (SLD) profile as a function of the coordinate $z$;
normal to the sample surface~\cite{Xiao-Lin2,Majkrzak7}. As the thickness of the sample is of the order of nanometer,
 multiple scattering is neglected in neutron specular reflectometry and the neutron specular reflection is accurately
 described by a one-dimensional Schr\"{o}dinger wave equation
\begin{equation} \label{eq1}
\left\{\partial_{z}^2+\left[q_{0}^2-4\pi\rho(z)\right]\right\}\psi(q,z)=0,
\end{equation}
where $\psi(q,z)$  is the neutron wave function and $q_{0}$ is the $z$-component of the incident neutron wave vector in vacuum.

Specular reflection is rigorously invertible, however, if both the modulus and the phase of the complex
 reflection coefficient, $r(q_{0})$, were known, a unique result for scattering length density, $\rho(z)$ would be determined~\cite{Masoudi8}.

An exact relation for the reflection coefficient $r(q_{0})$ can be derived from the transfer matrix which
 is a $2\times 2$ matrix with the elements $A(q_{0}), B(q_{0}), C(q_{0})$, and $D(q_{0})$ which are determined by solving
 the equation~(\ref{eq1}),  for a film with SLD $\rho(z)$~\cite{Penfold3,Majkrzak7}.

Correspondingly, for a sample with arbitrary surroundings, the solution of equation~(\ref{eq1}) in terms of the transfer matrix elements is represented by:
\begin{equation} \label{eq2}
\begin{pmatrix}
1 \\
\ri h \\
\end{pmatrix} t\re^{\ri hqL}=
\begin{pmatrix}
A & B \\
C & D \\
\end{pmatrix}
\begin{pmatrix}
1+r \\
\ri f(1-r) \\
\end{pmatrix},
\end{equation}
where $t(q)$ is the transmission coefficient and $f$ and $h$ are the refractive index of the fronting and backing medium
 having SLD values of $\rho_{f}$ and $\rho_{h}$ respectively. For non-vacuum surroundings,  $n=\Big(1-4\pi\frac{\rho_{n}}{q_{0}^2}\Big)^{1/2}$
where $n$ stands for $f$ and $h$. Correspondingly, for vacuum fronting or backing $n(q)= f=h=1$~\cite{Majkrzak1}.

The transfer matrix has a unit determinant and is unimodular;
\begin{equation} \label{eq3}
A(q_{0}) D(q_{0})-B(q_{0}) C(q_{0})=1.
\end{equation}
As the other property of the transfer matrix, reversing the potential, $\rho(z)\longrightarrow\rho(L-z)$, would cause the
 interchange of the diagonal elements $A$ and $D$ without having any effect on the off-diagonal elements, $B$ and $C$~\cite{Majkrzak1}.

By solving the equation~(\ref{eq1}) for the reflection coefficient, $r(q)$, we have:
\begin{equation} \label{eq4}
r=\frac{\left(f^2h^2B^2+f^2D^2\right)-\left(h^2A^2+C^2\right)-2\ri\left(fh^2AB+fCD\right)}
{\left(f^2h^2B^2+f^2D^2\right)+\left(h^2A^2+C^2\right)+2fh}\,.
\end{equation}
The reflectivity $R(q)=\vert r(q)\vert ^2$ can be represented in terms of the transfer matrix elements by introducing a new quantity $\Sigma(q)$:
\begin{equation} \label{eq5}
\Sigma(q)=2\frac{1+R^{fh}}{1-R^{fh}}= \left(f^2h^2B^2+f^2D^2\right)+\left(h^2A^2+C^2\right)
\end{equation}
where $R$ is the amplitude of the complex reflection coefficient and the superscript $fh$, denotes a sample with fronting
 and backing medium with refractive index $f$ and $h$, respectively. By introducing the following new quantities
\begin{align} \label{eq6}
&\alpha ^{fh}=\left(f^{-1}hA^2+f^{-1}h^{-1}C^2\right), \nonumber\\
&\beta ^{fh}=\left(fhB^2+fh^{-1}D^2\right), \nonumber\\
&\gamma^{fh}=hAB+h^{-1}CD.
\end{align}
Equations (\ref{eq4}) and (\ref{eq5}), are represented as
\begin{equation} \label{eq7}
r^{fh}=\frac{\beta^{fh}-\alpha^{fh}-2\ri\gamma^{fh}}{\beta^{fh}+\alpha^{fh}+2}\, ,
\end{equation}
\begin{equation} \label{eq8}
\Sigma(q)=\beta^{fh}+\alpha^{fh}.
\end{equation}
These two equations are the two key relations in reference method. Equation~(\ref{eq2})
shows that below the critical wave number $q_{\mathrm{c}}=4\pi {\rm Max}\left(\varrho_{f},\varrho_{h}\right)^{1/2}$,
the method is not applicable since the information of $r(q)$ is not available. However, the missing data below
the $q_\mathrm{c}$ could be accurately retrieved by extrapolation~\cite{Majkrzak1,Majkrzak7}. In section~4 we will introduce an
 algebraic method for determining the missing value below the critical~$q$.

\subsection{Reference layer method} \label{Reflayer}

Here we explain how one can find the complex reflection coefficient of any unknown non-absorptive layer by using some
reference layers. Suppose a known layer has been mediated between the fronting medium and the unknown layer. In this case,
 the transfer matrix of the whole sample (containing the known and unknown layers) can be expressed by the multiplication of
 the transfer matrix of each one as follows:
\begin{equation} \label{eq9}
\begin{pmatrix}
A & B \\
C & D \\
\end{pmatrix}=
\begin{pmatrix}
a & b \\
c & d \\
\end{pmatrix}
\begin{pmatrix}
w & x \\
y & z \\
\end{pmatrix},
\end{equation}
where $(a,\dots,d)$ refer to the matrix elements of the unknown part and $(w,\dots,z)$ present the matrix elements of the known part. Consequently we have
\begin{eqnarray}
 \label{eq10}
\Sigma=\left(h^2a^2+c^2\right)\left(w^2+f^2x^2\right)
+\left(h^2b^2+d^2\right)\left(y^2+f^2z^2\right)+2\left(h^2ab+cd\right)\left(wy+f^2xz\right).
\end{eqnarray}
Alternatively, the above formula can be denoted in compact notation in terms of three unknown real-valued parameters as
\begin{align}\label{eq11}
\Sigma(q)=2\frac{1+R^{fh}}{1-R^{fh}}= h^2\widetilde{\beta}^{ff}_{\mathrm K} \alpha^{hh}_{\mathrm U} + f^2\widetilde{\alpha}^{ff}_{\mathrm K} \beta^{hh}_{\mathrm U} +
 2fh\widetilde{\gamma}^{ff}_{\mathrm K} \gamma^{hh}_{\mathrm U}\, ,
\end{align}
where the subscripts $\mathrm K$ and $\mathrm U$ refer to the known and unknown parts of the sample and the tilde represents the parameters of the
 sample with mirror reversed potential $(A \longleftrightarrow D)$. $ff$ and $hh$ denote the same fronting and backing media having
the SLD values of $\rho_{f}$ and $\rho_{h}$, respectively.

In order to determine the reflection coefficient, $r(q)$, we consider three measurements of reflectivity for the three known reference layers.
Using equation~(\ref{eq6}) and (\ref{eq11}) for three different reference layers $(\rm K_{1}, \rm K_{2}, \rm K_{3})$, three equations will be obtained.
By solving these equations in matrix algebra as shown in equation~(\ref{eq12}), the parameters $(\alpha^{hh}_{\mathrm U}, \beta^{hh}_{\mathrm U}, \gamma^{hh}_{\mathrm U})$ are
determined which represent the $r^{hh}(q)$ for the unknown film in contact with a uniform  identical surroundings in front and back with $\rho_{h}$.

\begin{equation} \label{eq12}
\begin{pmatrix}
\alpha^{hh}_{\mathrm U}\\
\beta^{hh}_{\mathrm U}\\
\gamma^{hh}_{\mathrm U}
\end{pmatrix}={\bf M}
\begin{pmatrix}
\Sigma_{1}\\
\Sigma_{2}\\
\Sigma_{3}
\end{pmatrix}, \qquad
{\bf M}=\begin{pmatrix}
h^2\widetilde{\beta}^{ff}_{\mathrm K_{1}} & f^2\widetilde{\alpha}^{ff}_{\mathrm K_{1}} & 2fh\widetilde{\gamma}^{ff}_{\mathrm K_{1}}\\
h^2\widetilde{\beta}^{ff}_{\mathrm K_{2}} & f^2\widetilde{\alpha}^{ff}_{\mathrm K_{2}} & 2fh\widetilde{\gamma}^{ff}_{\mathrm K_{2}}\\
h^2\widetilde{\beta}^{ff}_{\mathrm K_{3}} & f^2\widetilde{\alpha}^{ff}_{\mathrm K_{3}} & 2fh\widetilde{\gamma}^{ff}_{\mathrm K_{3}}
\end{pmatrix}^{-1}.
\end{equation}
For $q>q_{\rm c}$ this gives us the reflection coefficient
\begin{equation} \label{eq13}
r_{\mathrm U}^{hh}=\frac{\beta_{\mathrm U}^{hh}-\alpha_{\mathrm U}^{hh}-2\ri\gamma_{\mathrm U}^{hh}}{\beta_{\mathrm U}^{hh}+\alpha_{\mathrm U}^{hh}+2}\, ,
\end{equation}
which is equivalent to the reflectivity of a ``free film'' defined by $\rho_{\mathrm{equiv}}(z)=\rho_{\mathrm U}(z)-\rho_{h}$. By inverting $r_{\mathrm U}^{hh}$ and
 shifting by a known constant value, the SLD of the free film can be retrieved~\cite{Majkrzak1,Majkrzak5}.

\subsection{Determining the phase below the critical wave number}
There are several ways for retrieving the missing data below the critical wave number such as extrapolation or using polarized neutron and
 a magnetic substrate. Here we introduce an algebraic method to retrieve the data below $q_{\rm c}$.

The transfer matrix for a layer having constant SLD in terms of wave number is represented by:
\begin{equation} \label{eq14}
{\bf T}=\begin{pmatrix}
\cos nq_{0}L    & \frac{1}{n}\sin nq_{0}L\\
-n\sin nq_{0}L & \cos nq_{0}L
\end{pmatrix},
\end{equation}
where $n=\sqrt{1-4\pi \varrho/q_{0}^2}$, $L$ is the thickness and $\varrho$ is the SLD of the sample.

Introducing several new parameters, $\varrho_{\rm eq}=\varrho-\varrho_{h}$, $q_{\rm eq}=\sqrt{q_{0}^2-4\pi \varrho_{h}}$ and
 $n_{\rm eq}=\sqrt{1-4\pi \varrho_{\rm eq}/q_{\rm eq}^2}$, for ${\bf T}$ we have
\begin{equation} \label{eq15}
{\bf T}=\begin{pmatrix}
\cos n_{\rm eq}q_{\rm eq}L    & \frac{1}{hn_{\rm eq}}\sin n_{\rm eq}q_{\rm eq}L\\
-hn_{\rm eq}\sin n_{\rm eq}q_{\rm eq}L & \cos n_{\rm eq}q_{\rm eq}L
\end{pmatrix}.
\end{equation}

By inserting the elements of transfer matrix in equation~(\ref{eq13}) and using equation equation~(\ref{eq7}), we have:
\begin{equation} \label{eq16}
r^{hh}(\varrho)\Big|_{q_{0}=\sqrt{4\pi \varrho_{h}}}=r^{11}(\varrho-\varrho_{h})\Big|_{q_{j}=\sqrt{q_{0}^2-4\pi \varrho_{h}}}\,.
\end{equation}
Equation~(\ref{eq16}) shows that knowing $r^{hh}(\varrho)$ for $q_{0}\geqslant\sqrt{4\pi \varrho_{h}}$, the reflection coefficient
 of a free sample having SLD value of $\varrho_{\rm eq}=\varrho-\varrho_{h}$, $r^{11}(\varrho-\varrho_{h})$, is known for
 specific wave number $q_{j}=\sqrt{q_{0}^2-4\pi \varrho_{h}}$  for $q_{j}\geqslant0$.

\subsection{Smooth variation of interface}\label{smoothness}

As we mentioned in section~\ref{Reflayer}, there is some smearing at boundaries in real samples and the SLD varies smoothly and
 continuously from one layer to another. The smooth variation of SLD at boundaries can be expressed by Error function as follows \cite{Majkrzak6}:
\begin{equation} \label{eq17}
\varrho(z)=\varrho_{1}(z)+\frac{\varrho_{2}(z)-\varrho_{1}(z)}{2}
\left[1+\mathrm{erf}\left(\frac{z-\Delta}{\sqrt{2}\sigma}\right)\right],
\end{equation}
where $\sigma$ is the smoothness factor. In other word, $\sigma$ is the thickness of the smoothly varying area across the
 interface. $\Delta$ is also the mean position of the interface or the turning point of the Error function. Since the reference
 layer method is based on discontinuity of the interfacial potential, it is obvious that the effect of smoothness at boundaries
 would affect the reconstruction procedure. In the next section we have investigated these effects numerically.

\section{Numerical example}\label{numerical}

\begin{figure}[!t]
\centering{\includegraphics[width=8.8cm]{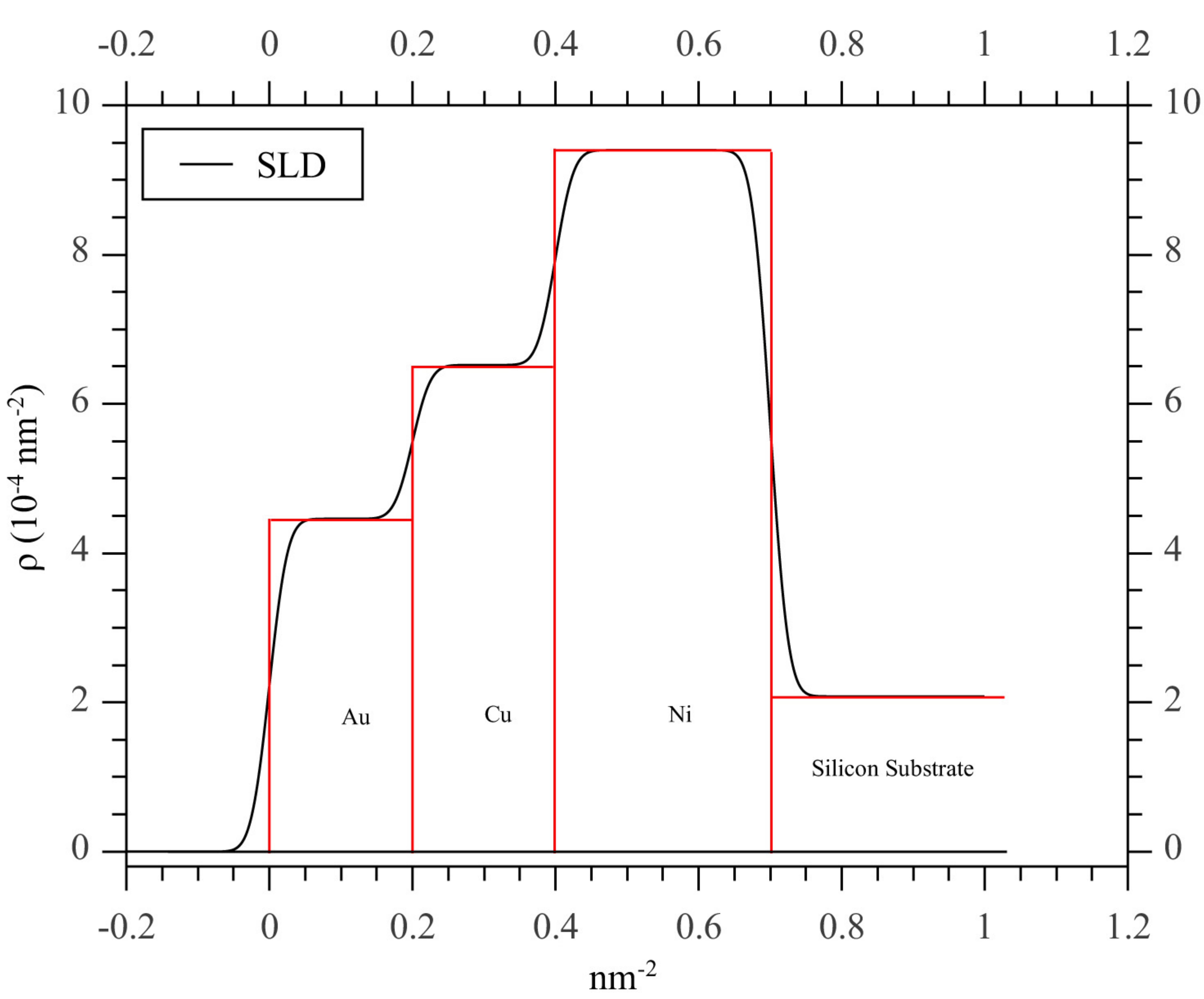}} \caption{A known reference layer over the Cu-Ni sample with smooth potential at boundaries. The red lines demonstrate the ideal sample with non-smooth boundaries.}
\label{fig2}
\end{figure}
\begin{figure}[!b]
\centering{\includegraphics[width=8.8cm]{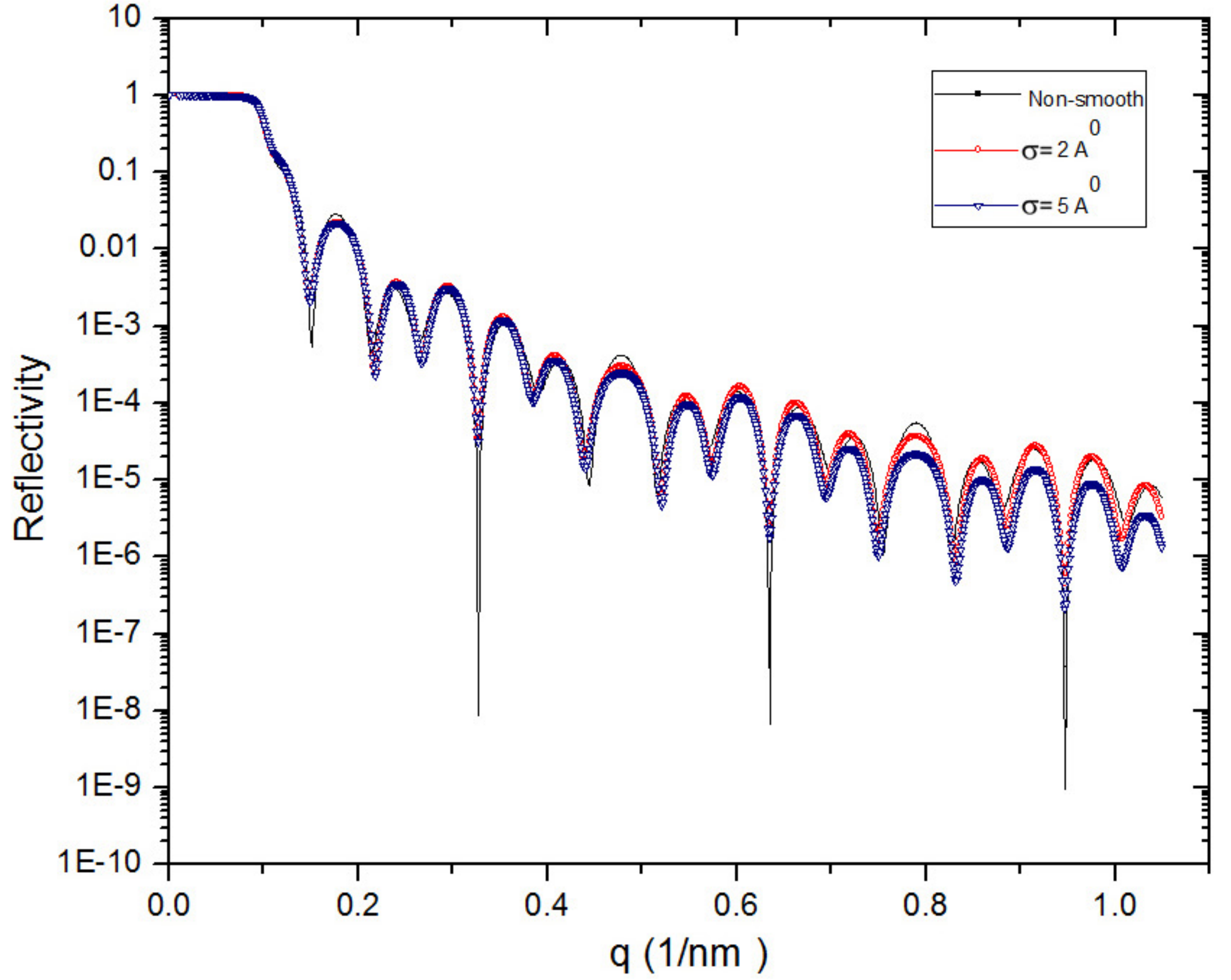}} \caption{(Color online) Reflectivity from the sample of figure~\ref{fig2} for three cases: $\sigma=0,\ 0.2$ and 0.5~nm.}
\label{fig3}
\end{figure}

As a realistic example to study the stability of the reference method by using a smooth potential, we consider a sample with 20~nm thick Copper on 30~nm Nickel having a constant SLD of $6.52\times10^{-4}$~nm$^{-2}$ and $9.4\times10^{-4}$~nm$^{-2}$ as the unknown layer; and three separate reference layers with 20~nm Au, 15~nm Cr and 10~nm Co having a constant SLD value of $4.46, 3.03$ and 2.23 ($\times10^{-4}$~nm$^{-2}$), respectively. Figure~\ref{fig2} shows this configuration for the gold layer as reference. Figure~\ref{fig3} also shows the reflectivity of the sample of figure~\ref{fig2} for three difference smoothness factors. The smearing of potential at boundaries is taken into account by using $\sigma =0,\ 0.2$ and 0.5~nm. The effects of smoothness on output results are clear in the figure, particularly at large wave numbers, while the data for small wave numbers truly correspond with none-smooth data.

\begin{figure}[h]
\centering{\includegraphics[width=8.8cm]{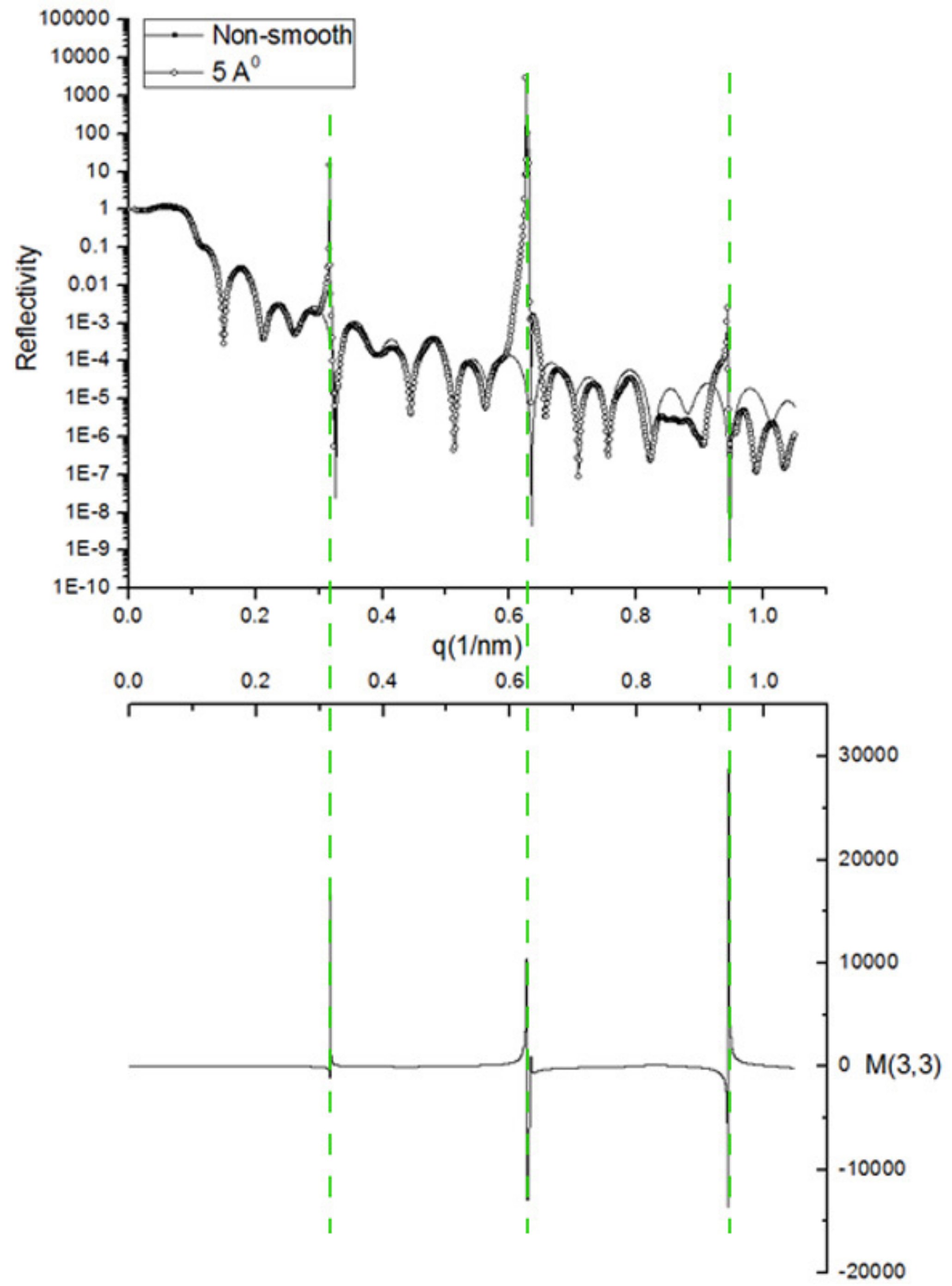}} \caption{Reflectivity of the unknown sample of figure~\ref{fig2} which
 is reconstructed by using three reference layers: 20~nm gold, 15~nm Cr and 10~nm Co (Circled curve is for $\sigma=0.5$~nm and
 solid line is for the ideal sample, $\sigma=0$).  ${\bf M}(3,3)$ element of the ${\bf M}$ matrix, equation~(\ref{eq12}), is plotted for better
 understanding of the origin of reflectivity noises.}
\label{fig4}
\end{figure}

\begin{figure}[h]
\centering{\includegraphics[width=8.5cm]{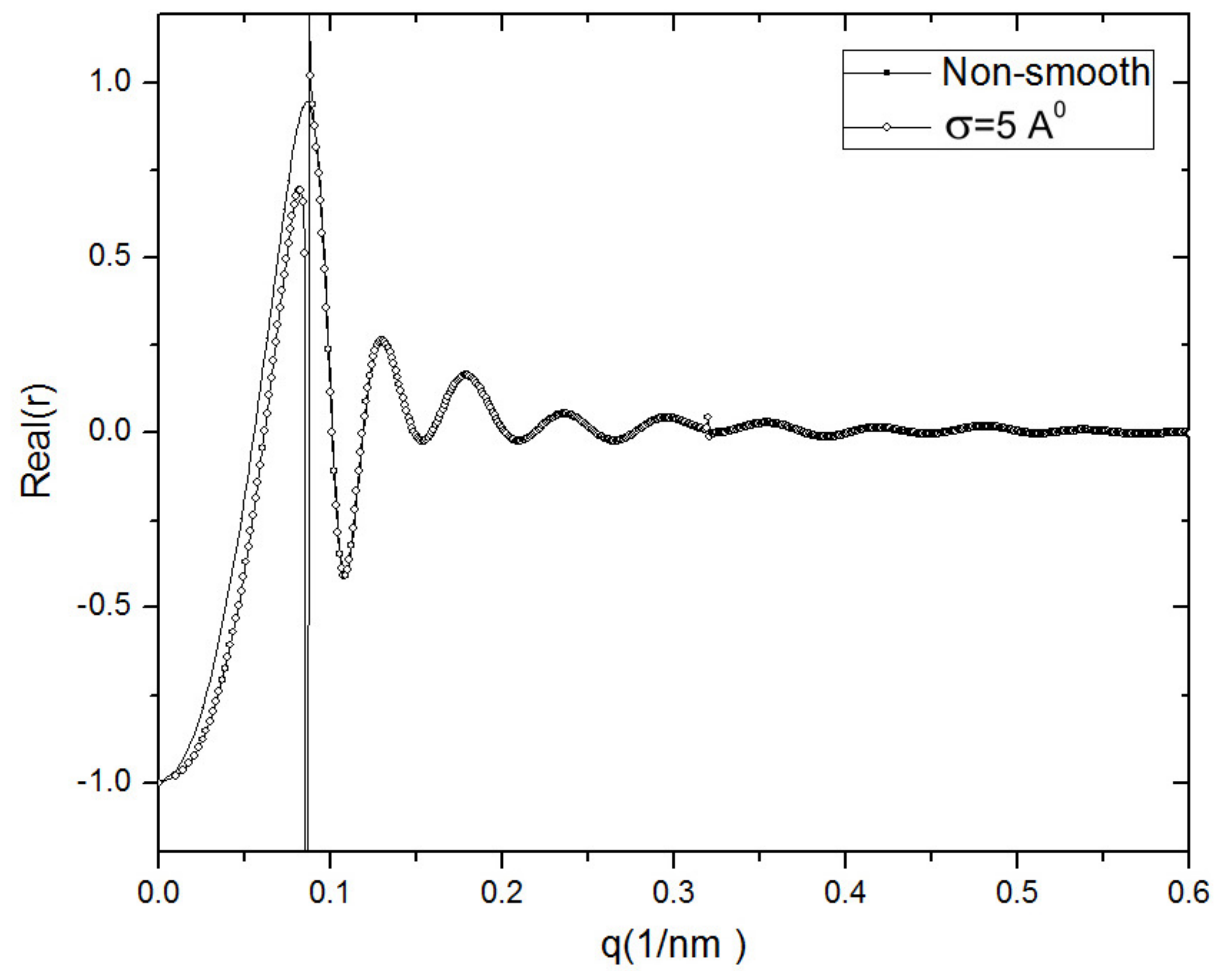}} (a)%
\\
\includegraphics[width=8.5cm]{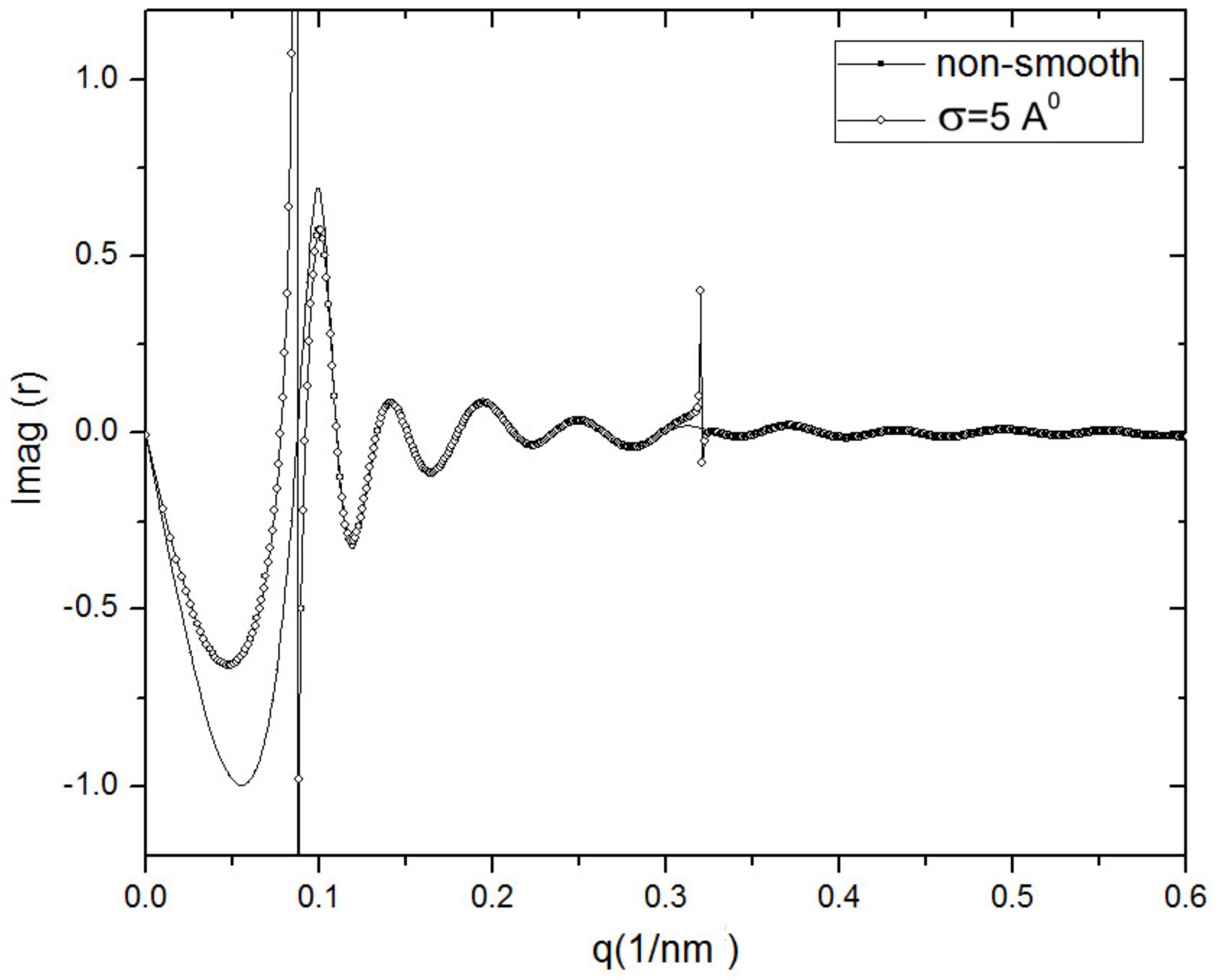} (b)%
\caption{The reflection coefficient of the unknown part of the sample of figure~\ref{fig2} for two different $\sigma=0$, 5~\AA \
 by three reference  layers as; 20~nm Au, 15~nm Cr and 10~nm Co.  (a) Real and (b), imaginary parts of the reflection coefficient.}
\label{fig5}
\end{figure}

Reflectivity of the unknown part of the sample with smooth varying SLD at the interface with $\sigma=5$~\AA \ is illustrated
in figure~\ref{fig4} (upper curve). Similar to figure~\ref{fig3}, the effects of smoothness are more obvious at large wave numbers.
However, consideration of smoothness at interface would cause some abrupt noises at certain values of wave numbers such
as: $q\simeq0.3$ and 0.6~nm$^{-1}$. The noises also appear for the real and imaginary data of reflection
coefficient (figure~\ref{fig5}~(a), (b)). These noises are due to the abrupt changes in some of the elements of
the ${\bf M}$ matrix, equation~(\ref{eq12}), in the reference method. The intensity and distribution of these noises along the different
 ranges of wave numbers are extremely relevant to the smoothness factor, thickness of the reference layers and the range
 of neutron wave numbers. In order to  better understand the origin of these noises, we have plotted the ${\bf M}(3,3)$ element of
 the ${\bf M}$ matrix in figure~\ref{fig4} (lower curve). The result shows that ${\bf M}(3,3)$ has some noises in $q$ values, exactly  as for
 the reflectivity data. Such severe noises in output results would cause a complete loss of reflection coefficient data in
 those $q$ values. As we mentioned in section~\ref{Theory}, the data of reflection coefficient in the whole range of wave numbers,
 even bellow the critical wave number, are needed to retrieve the SLD of the sample. Unfortunately, the noises are distributed
over different values of wave numbers ranging from small to large $q$'s. This deficiency makes the data useless in retrieving the SLD of the sample. Here we are going to propose a method for removing or reducing these noises.

\begin{figure}[h]
\centering{\includegraphics[width=8.5cm]{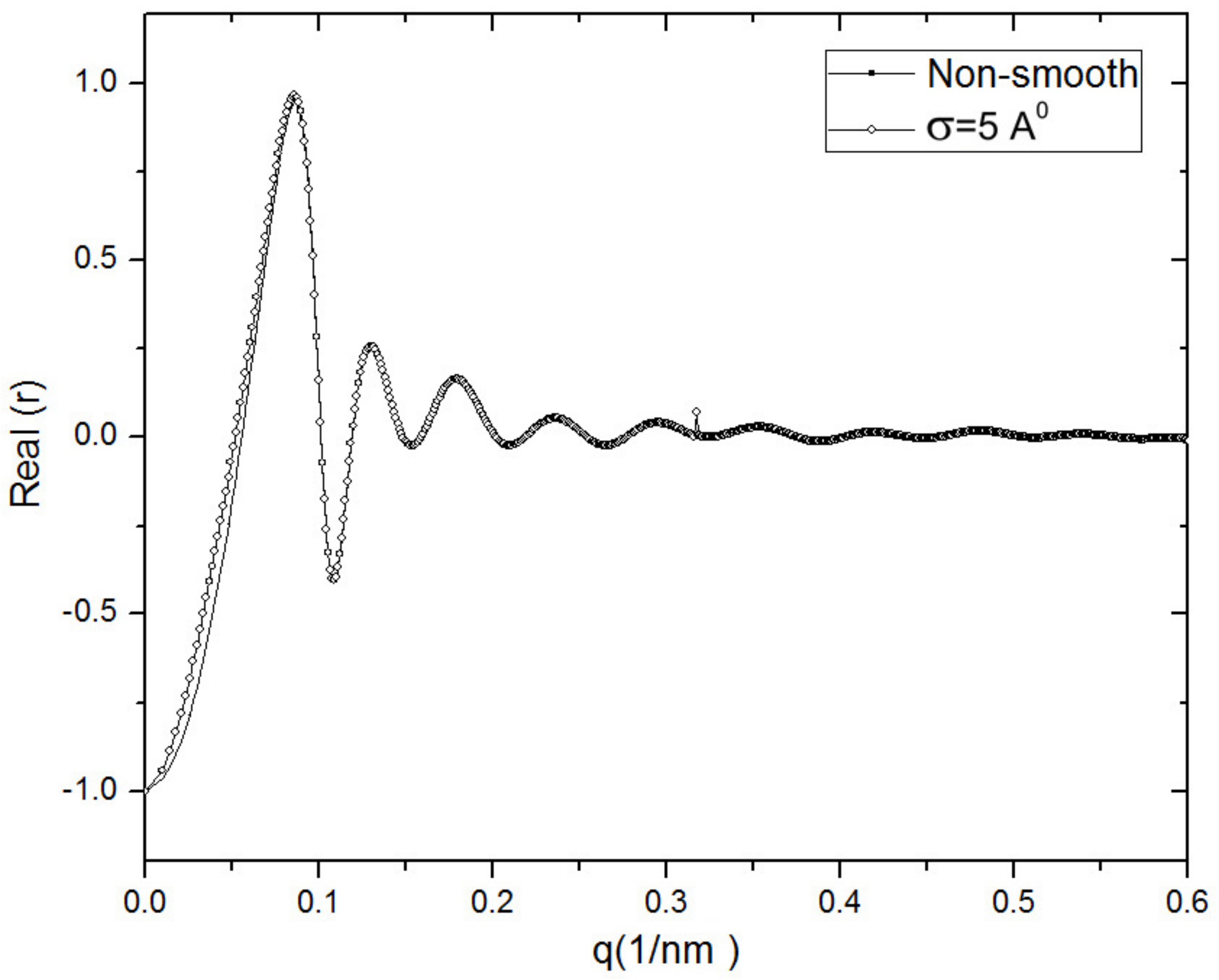}}  \centering (a) \\
\includegraphics[width=8.5cm]{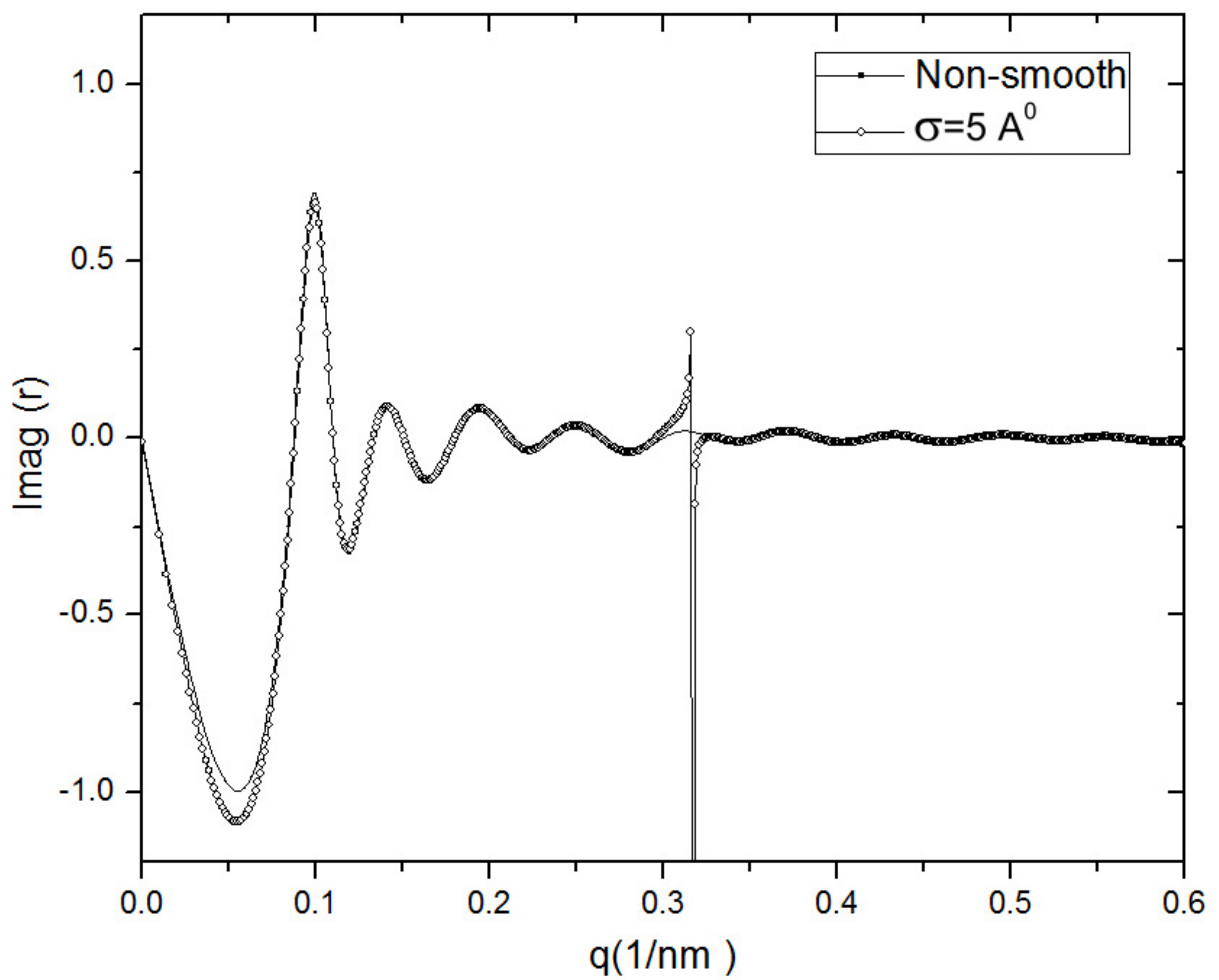} \centering (b)
\caption{The reflection coefficient of the unknown part of the sample of figure~\ref{fig2} for $\sigma=0$, 5~\AA. The three reference
 layers are: 15~nm Au, 20~nm Cr and 10~nm Co. Due to the proper selection of reference layers, the noise is removed in comparison
 with the result of figure~\ref{fig4}.  (a) Real and  (b), imaginary parts of the reflection coefficient.}
\label{fig6}
\end{figure}

The stability of the method to a great extent depends on the ${\bf M}$ matrix which is a function of the known part of the sample.
The best result for the reflection coefficient of the unknown part of the sample comes out of the equation~(\ref{eq12}) in the case where the behavior of the ${\bf M}$ matrix is smooth. Our investigation shows that not every arbitrary reference layer can lead to a smooth result for the ${\bf M}$ matrix. Since we can control the choice of the reference layers, we can run a simulation to find the best layers for which the ${\bf M}$ matrix behaves smoothly and without noise. By choosing the best reference layer, we can retrieve reliable results for the  unknown part of the sample. Those reference layers which are well behaved can be used for experimental implementation  of the reflectometry experiments. As we mentioned before, the noise strongly depends on the thickness of the reference layers, and the choice of  a proper thickness for the reference films would enhance the accuracy of the output results. In our example, the numerical  calculations show that the best result with the smallest noise level would be obtained for 15, 10 and 20~nm thicknesses for Au, Co and Cr, respectively.  The purified curves of real and  imaginary parts of reflection coefficient in this case are shown in figures~\ref{fig6}~(a), (b). As it is clear in the  figure, we have much better results with less noise in comparison with figure~\ref{fig5}.

Thereupon, the choice of an appropriate thickness for the reference layers would completely remove the noise or displace
it to large wave numbers, where it would cause no ambiguity in retrieving the SLD profile of the sample. The rest
 of the noises can be removed by extrapolation. By using these corrected data as input for some useful codes such as
 the one which is developed by Sacks~\cite{Sacks9,Sabatier10,Sacks11} based on the Gel'fand-Levitan integral
equation~\cite{Majkrzak1,Xiao-Lin2,Penfold3}, the scattering length density of the sample is retrieved. The data of
reflection coefficient up to $q=0.6$ are sufficient to retrieve the SLD by Sacks code and we can neglect the data of large
wave numbers. Hence, after purification of data, any remaining noise in the range of large wave numbers can be neglected.

\begin{figure}[!t]
\centering{\includegraphics[width=8.cm]{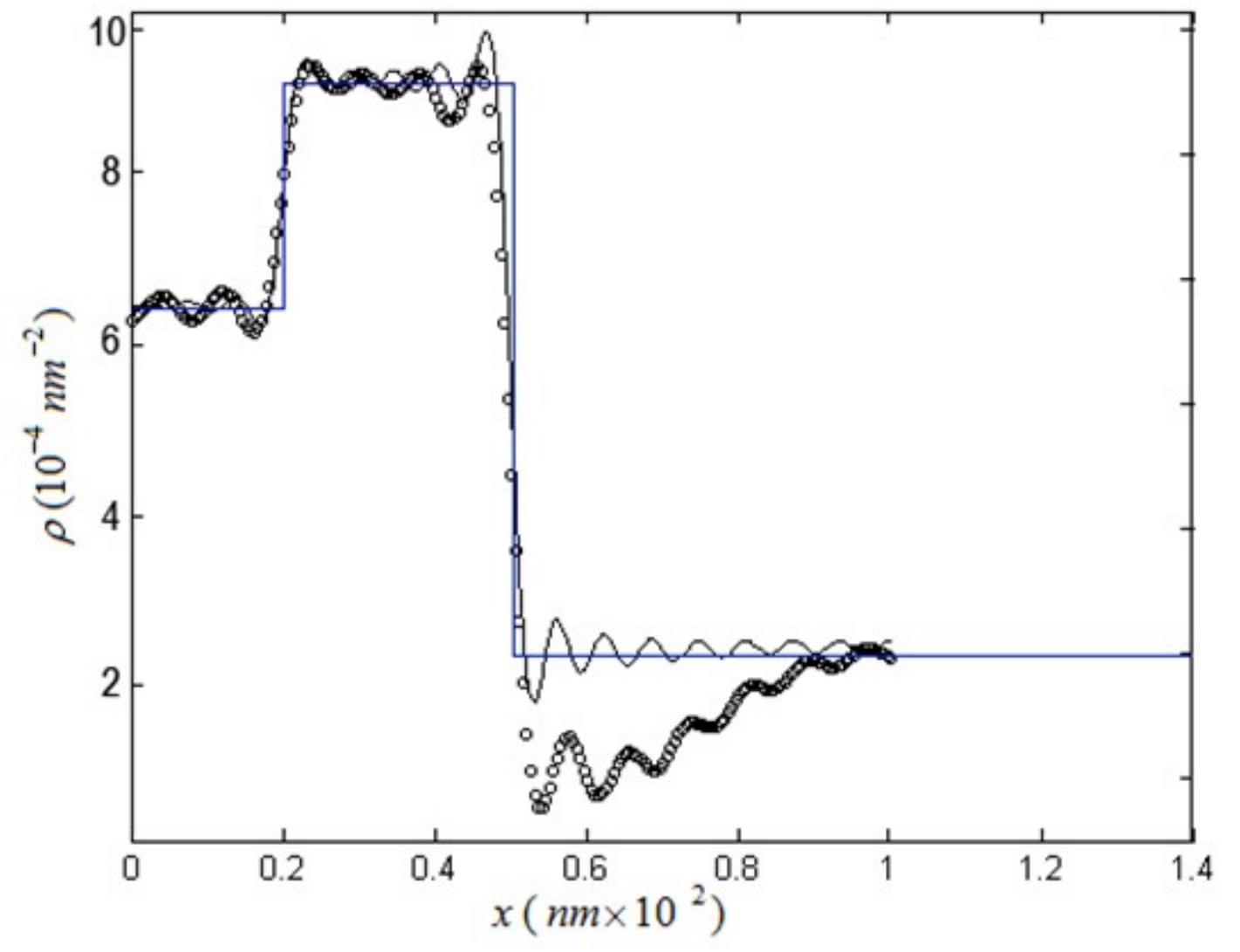}} \caption{The reconstructed SLD of the unknown sample
shown in figure~\ref{fig2}, surrounded by identical medium on both sides ($\varrho_{hh}$), obtained by inversion
of the extracted values of the reflection coefficient in the momentum range $0 \leqslant q \leqslant0.6$~nm$^{-1}$. The non-smooth
profile is shown by thin solid line. The circled curve represents the SLD of the sample in the presence of smoothness.}
\label{fig7}
\end{figure}

The SLD profile corresponding to $r^{hh}(q)$ was retrieved by using the real and imaginary parts of the reflection
coefficient at the presence of the smoothness. As it is shown in figure~\ref{fig7}, the method of reference layers is stable
at the presence of smoothness and the SLD corresponding to $r^{hh}(q)$ is truly reconstructed. The circled curve
depicts the retrieved SLD profile of the sample. The retrieved thicknesses corresponds well with the thickness of the sample.

\begin{figure}[!b]
\centering{\includegraphics[width=8.cm]{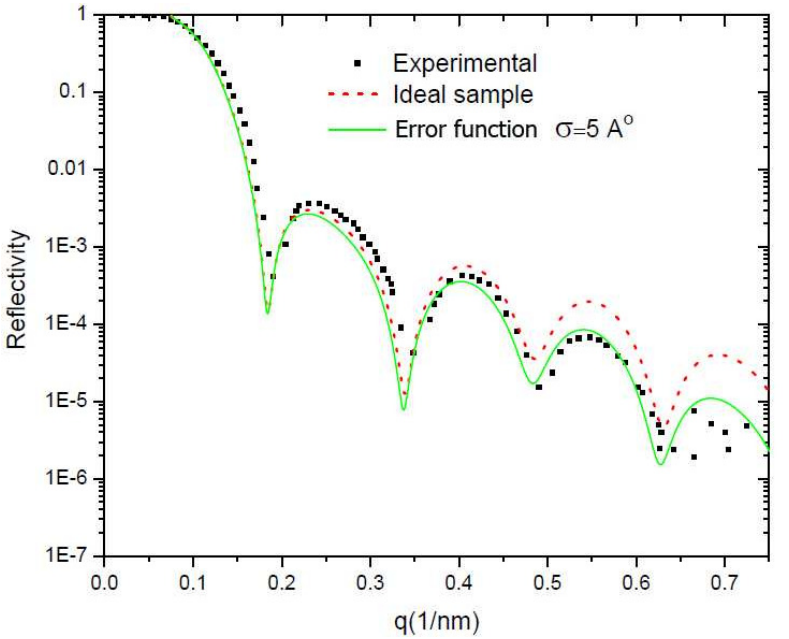}} \caption{(Color online) Experimental reflectivity curve in contrast to the
simulated reflectivity at the presence of smoothness of boundaries and the ideal sample. The green solid line demonstrates
the reflectivity at the presence of smoothness and better corresponds with the experimental curve.}
\label{fig8}
\end{figure}

Consideration of smoothness of interfaces makes the output results more corresponding to experimental data. In
order to verify the efficiency of the method, the reflectivity curves were plotted in figure~\ref{fig8} for a sample
from~\cite{Majkrzak6}, Al$_{2}$O$_{3}$(6.7~nm)/FeCo(20~nm)/GaAS, for either of smooth (green solid line) and non-smooth
 boundaries (red dashed line), in contrast to the experimental reflectivity curve.
As it is shown in the figure, the reflectivity curve in the presence of smoothness of interfaces better corresponds with
the experimental reflectivity curve while the simulated reflectivity for an ideal sample does not absolutely correspond with
the experimental curve, particularly at large wave numbers. The results verify that consideration of the smoothness of interfaces is of highest importance.

\section{Conclusion}
In most of the simulation methods of neutron reflectometry, the SLD of adjacent layers is supposed to be discontinuous and sharp. Since in real samples there is observed some smearing at boundaries, consideration of smoothness would effect the output results and the retrieved data.

In this paper by considering a certain type of Error function which denotes the smooth variation of SLD across the interface
 and implementing it into the reference method, the effects of smoothness on determining the phase and retrieving the SLD
 profile of the sample were investigated. Due to consideration of smoothness of interfaces, several abrupt noises existed
 in the output, such as reflectivity, real and imaginary parts of reflection amplitude. It was shown that these noises
 are due to the abrupt change in the elements of ${\bf M}$ matrix of equation~(\ref{eq12}). We proposed to choose a proper thicknesses
 for the reference layers, which would completely remove the noise or shifted it to the large wave numbers $(q>0.6)$, where the
 data of reflectivity are not needed in order to retrieve the SLD and we can easily neglect them. Consequently, the SLD of the sample
 was retrieved by using the purified data of the reflection coefficient.

In order to make sure that the method is highly reliable, we also compared the simulated reflectivity with the experimental
 curve for a sample. Astoundingly, the results show that consideration of smooth variation of SLD at the interface would
 make the measured reflectivity to better correspond with the experimental data. The results also certify that the
 reference method is stable in the presence of smooth interfaces.


\ukrainianpart

\title{Дослідження  впливу гладкості міжфазових границь \\ на стійкість
зондуючих наномасштабних тонких плівок \\ за допомогою нейтронної рефлектометрії}

\author{С.С. Джаромі, С.Ф. Масуді}

\address{Фізичний факультет, Технологічний університет ім. K.Н. Тусі, а/с
15875-4416, Тегеран, Іран}

\makeukrtitle

\begin{abstract}
\tolerance=3000%
Більшість методів рефлектометрії, що використовуються для
визначення фази комплексного коефіцієнту відбивання такі як еталонний
метод і зміна прилеглого середовища грунтуються на розв'язку рівняння
Шредінгера з використанням розривного і сходинкоподібного оптичного
потенціалу розсіювання. Проте, під час процесу напорошування для
підготовки реального зразка два сусідні шари змішуються  і міжфазова
границя може не бути розривною і чіткою. Розмивання сусідніх шарів
при міжфазовій границі (гладкість інтерфейсу), може мати вплив на
відбивання, фазу коефіцієнта відбивання і перебудову густини довжини
розсіювання зразка. В цій статті ми дослідили стійкість еталонного
методу у присутності гладких міжфазових
границь. Гладкість міжфазових границь розглядається, використовуючи
неперервну функцію потенціалу розсіювання. Ми також запропонували
метод для отримання найбільш надійного результату, який відновлює
густину довжини розсіювання зразка.

\keywords нейтронна рефлектометрія, тонкі плівки, гладкий потенціал,
еталонний метод

\end{abstract}


\begin{thebibliography}{99}

\bibitem{Majkrzak1}  Majkrzak C.F.,  Berk N.F., Perez-Salas U.A., Langmuir, 2003, {\bf 19}, 7796;
\doi{10.1021/la0341254}.

\bibitem{Xiao-Lin2}  Zhou X.-L.,  Chen S.-H., Phys. Rep., 1995, {\bf 257}, 223; \doi{10.1016/0370-1573(94)00110-O}.

\bibitem{Penfold3}  Penfold J.,  Thomas R.K., J. Phys.: Condens. Matter, 1990, {\bf 2}, 1369; \doi{10.1088/0953-8984/2/6/001}.

\bibitem{Kasper4}  Kasper J.,  Leeb H.,  Lipperheide R., Phys. Rev. Lett., 1998, {\bf 80}, 2614;  \doi{10.1103/PhysRevLett.80.2614}.

\bibitem{Masoudi8} Masoudi S.F.,  Pazirandeh A., Physica B, 2005, {\bf 356}, 21; \doi{10.1016/j.physb.2004.10.038}.

\bibitem{jahromi} Jahromi S.S.,  Masoudi S.F., Appl. Phys. A, 2010, {\bf 99}, 255; \doi{10.1007/s00339-009-5515-5}.

\bibitem{Majkrzak7}  Majkrzak C.F.,  Berk N.F., Physica B, 2003, {\bf 336}, 27; \doi{10.1016/S0921-4526(03)00266-7}.

\bibitem{Majkrzak5}  Majkrzak C.F.,  Berk N.F., Physica B, 1996, {\bf 221}, 520; \doi{10.1016/0921-4526(95)00974-4}.

\bibitem{Majkrzak6} Fitzsimmons M.R., Majkrzak C.F., Application of polarized neutron reflectometry to studies of artificially structured magnetic materials. In: Modern Techniques for Characterizing Magnetic Materials, edited by Yimei Zhu, Springer, New York, 2005; \doi{10.1007/0-387-23395-4_3}.

\bibitem{Sacks9} Klibanov M.V., Sacks P.E., J. Comput. Phys., 1994, {\bf 112}, 273; \doi{10.1006/jcph.1994.1099}.

\bibitem{Sabatier10}  Chadan K.,  Sabatier P.C., Inverse Problem in Quantum Scattering Theory, 2nd edn, Springer, New York, 1989.

\bibitem{Sacks11}  Aktosun T.,  Sacks P., Inverse Prob., 1998, {\bf 14}, 211; \doi{10.1088/0266-5611/14/2/001};\\
%
Sacks P., Aktosun~T., SIAM J. Appl. Math., 2000, {\bf 60}, 1340; \doi{10.1137/S0036139999355588};\\
%
Aktosun T.,  Sacks P.,  Inverse Prob., 2000, {\bf 16}, 821; \doi{10.1088/0266-5611/16/3/317}.


\end{thebibliography}
\end{document}